\documentclass[floatfix,notitlepage,twocolumn, superscriptaddress,longbibliography,prl]{revtex4-2}
\usepackage{graphicx}
\usepackage{amsmath}
\usepackage{amsfonts}
\usepackage{amssymb}
\usepackage{bm}
\usepackage{comment}
\usepackage{color}
\usepackage[breaklinks]{hyperref}
\usepackage[normalem]{ulem}
\usepackage{braket}
\hypersetup{
    citecolor = blue,
    colorlinks = true,
    urlcolor = blue
}

\def\d{{\partial}}
\def\s{{\sigma}}

\def\k{{ {\bm k} }}

\def\A{{ {\bm A} }}

\def\S{{ {\bm S} }}
\def\0{{ {\bm 0} }}
\def\w{{\omega}}
\def\a{{\alpha}}
\def\b{{\beta}}
\def\g{{\gamma}}

\def\ve{{\varepsilon}}
\def\r{{ {\bm r} }}
\def\rK{{ {\rm K} }}
\def\rR{{ {\rm R} }}
\def\rA{{ {\rm A} }}

\def\cV{{ {\cal V} }}

\def\hGK{{ \hat{G}^\rK }}
\def\hGR{{ \hat{G}^\rR }}
\def\hGA{{ \hat{G}^\rA }}
\def\hX{{ \hat{X} }}
\def\hG{{ \hat{\Gamma} }}
\def\hL{{ \hat{\Lambda} }}

\def\GRmGA{{ (G^\rR - G^\rA) }}

\usepackage{xcolor}

\allowdisplaybreaks[4]

\begin{document} 
\title{Probing Fermi-surface spin-textures via the nonlinear Shubnikov-de Haas effect}
\author{Kazuki Nakazawa}\affiliation{RIKEN Center for Quantum Computing (RQC), 2-1 Hirosawa, Wako, Saitama 351-0198, Japan}
\author{Henry F. Legg}\affiliation{SUPA, School of Physics and Astronomy, University of St Andrews, North Haugh, St Andrews, KY16 9SS, United Kingdom}
\affiliation{Department of Physics, University of Basel, Klingelbergstrasse 82, 4056 Basel, Switzerland}

\author{Renato M. A. Dantas}\affiliation{Department of Physics, University of Basel, Klingelbergstrasse 82, 4056 Basel, Switzerland}

\author{Jelena Klinovaja}\affiliation{Department of Physics, University of Basel, Klingelbergstrasse 82, 4056 Basel, Switzerland}

\author{Daniel Loss} \affiliation{Department of Physics, University of Basel, Klingelbergstrasse 82, 4056 Basel, Switzerland}
\affiliation{RIKEN Center for Quantum Computing (RQC), 2-1 Hirosawa, Wako, Saitama 351-0198, Japan}
\affiliation{Physics Department, King Fahd University of Petroleum and Minerals, 31261, Dhahran, Saudi Arabia}
\affiliation{Center for Advanced Quantum Computing, KFUPM, Dhahran, Saudi Arabia}
\affiliation{RDIA Chair in Quantum Computing}
\date{\today}

\begin{abstract}
The coupling of spin and electronic degrees of freedom via the spin-orbit interaction (SOI) is an essential ingredient for many proposed future technologies. However, probing the strength and nature of SOI is a significant challenge, especially in heterostructures. Here, we consider the nonlinear Shubnikov–de Haas (NSdH) effect, a quantum oscillatory effect that occurs under conditions similar to those of the well-known SdH effect, but is second order in the applied electric field. We demonstrate that, unlike its linear counterpart, the NSdH effect is highly sensitive to the spin textures that arise from SOI. We show that the relative phases and beating patterns of the nonlinear oscillations provide additional information that discriminates linear- and cubic-Rashba-dominated regimes in the models considered here. Our results establish NSdH as a complementary phase-resolved probe of SOI, offering a new framework for characterizing materials relevant to topology, spintronics, and solid-state quantum information technologies.
\end{abstract}

\maketitle

{\it Introduction}.---The spin--orbit interaction (SOI) is a cornerstone of modern condensed matter physics, driving key fields such as spintronics~\cite{ZFS2004,BP2010,Hirohata2020}, topological matter~\cite{AMV2018}, and quantum information processing~\cite{LD1998,HKPTV2007,KL2013,Scappucci2021,Hendrickx2020a,Hendrickx2020b,Hendrickx2021}, while also underpinning fundamental phenomena including the anomalous and spin Hall effects~\cite{KL1954,Nagaosa2010a,DP1971PLA,DP1971JETP,MNZ2003,Sinova2004,CL2005,CL2009,DML2009}, the Dzyaloshinskii–Moriya interaction~\cite{Dzyaloshinsky1958,Moriya1960}, and topological electronic band structures~\cite{KM2005,FK2007,HK2010,AMV2018}. The ability to electrically tune SOI in low-dimensional systems~\cite{Nitta1997,KTL2011,KRL2018,ABBKL2022,ABKL2022,Gao2020,VBPN2018,MVN2021,Bellentani2021,Piot2022,BHL2021,BL2022}, such as semiconductor nanostructures, is particularly important, enabling benefits such as longer spin-relaxation lengths~\cite{MC2000,Sasaki2014,Kunihashi2016}. For instance, the combination of strong SOI with magnetic fields and tailored device geometries can be exploited to suppress noise and improve spin qubit performance~\cite{Piot2022,BHL2021,BL2022}. The ability to characterize SOI and corresponding Fermi-surface spin textures is therefore of critical importance for future technologies.

A traditional method to detect features of a Fermi surface is the Shubnikov–de Haas (SdH) effect; the formation of Landau levels (LLs) in a strong magnetic field leads to sharp modulations in the electronic density of states, which in turn result in oscillations of the longitudinal conductivity perpendicular to the magnetic field~\cite{SdH1,SdH2,LK1956,LK1958,Shoenberg1984,Candido2023,YF2024,Zhang2024,CSES2025}.
The SdH oscillations are periodic in the out-of-plane inverse magnetic field $1/B_z$, with a frequency $F_i$ proportional to the extremal Fermi-surface area $S_i$ perpendicular to the magnetic field. This relationship between frequency and area means that SdH oscillations can map out the Fermi-surface of a solid~\cite{Shoenberg1984}. Despite the development of other powerful techniques such as angle-resolved photoemission spectroscopy, the SdH effect remains a common method for probing Fermi surfaces. For an SOI-split Fermi surface with areas $S_1$ and $S_2$ and frequencies $F_1$ and $F_2$, the oscillatory longitudinal conductivity may be written as $\sigma^{\rm osc}_{xx}\sim a_1\cos(2\pi F_1/B_z+\phi^{\rm L}_1)+a_2\cos(2\pi F_2/B_z+\phi^{\rm L}_2)$, where $\phi_i^{\rm L}$ is the phase offset of the $i$th orbit. In the regime considered here, $\phi^{\rm L}_1-\phi^{\rm L}_2\simeq0$, so their interference produces beating from which the SOI strength can be inferred~\cite{BR1984,Nitta1997,TA2002,AGT2005,Candido2023,Dettwiler2017,Beukman2017}, while providing only indirect access to the spin texture.
 
Recently, nonlinear charge transport~\cite{Sipe2000,GYN2014,SF2015,TSM2018,PMOM,JL,MP,DLX2021,Li2021,DWSO2021,NKM2022,Legg2022,OK2022,DLBLK,GCC2023,AXC2023,DLACA2023,Iwakiri2023,YNY2023,NYY2024,NYY2025,NLKL2024,Thenapparambil2026,Shibata2026} has emerged as a direct probe of SOI without requiring spin-current detection~\cite{Ideue2017,Li2021,DLBLK,NLKL2024}, with microscopic theories highlighting band geometry, spin texture, and orbital magnetic moments~\cite{NLKL2024,PMOM,MP,OK2022,MN,YNY2023,Shibata2026}. Previous nonoscillatory approaches characterize SOI through nonlinear response coefficients~\cite{Ideue2017,DLBLK,NLKL2024}. At finite $B_z$, however, nonlinear transport is nontrivial because the relevant eigenstates are field-dependent LL states rather than the momentum-space spinors. 

We therefore formulate the second-order conductivity directly in a LL basis, providing a unified framework from the weak-$B_z$ Fermi-surface regime to the more strongly quantized regime, and apply it to Ge 2DHGs with linear and cubic SOI~\cite{Scappucci2021,DLBLK,NLKL2024,XGLL2021}. The resulting nonlinear Shubnikov--de Haas (NSdH) effect provides a phase-resolved probe of SOI-induced spin textures. In the weak-$B_z$ intraband regime, where the two spin-split Fermi contours can be treated as distinct semiclassical orbits, a branch-resolved analysis gives opposite leading nonlinear weights for the two branches (see SM~\cite{com1}). Accordingly, for the models considered here, $\sigma^{\rm osc}_{xii}\sim b_1\cos(2\pi F_1/B_z+\phi^{\rm NL}_{1,i})+b_2\cos(2\pi F_2/B_z+\phi^{\rm NL}_{2,i})$, with $b_{1,2}>0$ and $\phi^{\rm NL}_{1,i}-\phi^{\rm NL}_{2,i}\simeq\pi$. Thus, unlike the approximately in-phase linear SdH contributions, the nonlinear contributions are approximately out of phase. At larger $B_z$, the branch-resolved interpretation becomes less accurate and the effective phase can deviate from $\pi$. These relative phases, including those between NSdH tensor components, provide spin-texture and SOI information beyond linear SdH; here, \lq\lq enhanced sensitivity'' refers to this phase information rather than to weaker thermal or disorder damping. 
 
{\it Formulation}.---We start with the derivation of the second-order response current to the electric field in the LL basis. In the Keldysh formalism, the statistical average of the current density is given by
\begin{align}
j_i \equiv
\langle \hat{j}_i \rangle
=
-\frac{i}{2}
\int \frac{d\ve}{2\pi}
{\rm tr}\left[
\hat{j}_i \hat{G}^{\rK}
\right],
\end{align}
where $\hat{G}^{\rK}$ is the Keldysh Green function, $\hat{j}_i=-\delta\hat{H}/\delta A_i$ is the $i^{\rm th}$ component of the electric current density operator, and
${\rm tr}$ denotes the trace over all degrees of freedom. To isolate the effect of the electric field, we separate the vector potential into space- and time-dependent parts, $\A(\r,t)\equiv\A'(\r)+\A''(t)$, and introduce the total and static kinetic momenta, $\hat{\bm\Pi} \equiv -i\hbar\bm\nabla-q\A(\r,t)$, $\hat{\bm\pi} \equiv -i\hbar\bm\nabla-q\A'(\r)$, respectively. Here, $q$ denotes the carrier charge. 
We define the static magnetic field $\bm B=\bm\nabla\times\A'(\r)$ and the spatially uniform electric field $\bm E=-\partial_t\A''(t)$. To calculate the response to $\bm E$, we expand the Hamiltonian in powers of the uniform vector potential as $\hat H(\hat{\bm\Pi}) = \hat H(\hat{\bm\pi}) + \hat\Gamma_i A_i''(t) + \hat\Gamma_{ij} A_i''(t)A_j''(t)+\cdots$, where $\hat\Gamma_i \equiv (\partial\hat H / \partial A_i'' )_{\A''=0}$ and $\hat\Gamma_{ij} \equiv \frac{1}{2} ( \partial^2\hat H / \partial A_i''\partial A_j'' )_{\A''=0}$. Repeated indices are summed over. Further details 
are given in the Supplemental Material  (SM)~\cite{com1}. Hereafter, we focus on the second-order response to the electric field. We first obtain the second-order ac current $j_i^{(2)}(\w_1,\w_2)$ at finite injected-field frequencies $\w_1$ and $\w_2$, and subsequently take the dc limit, retaining the nondivergent (regular) part~\cite{Kubo1957,MP,NLKL2024,com1}.
\begin{figure*}[t]
\includegraphics[width=180mm]{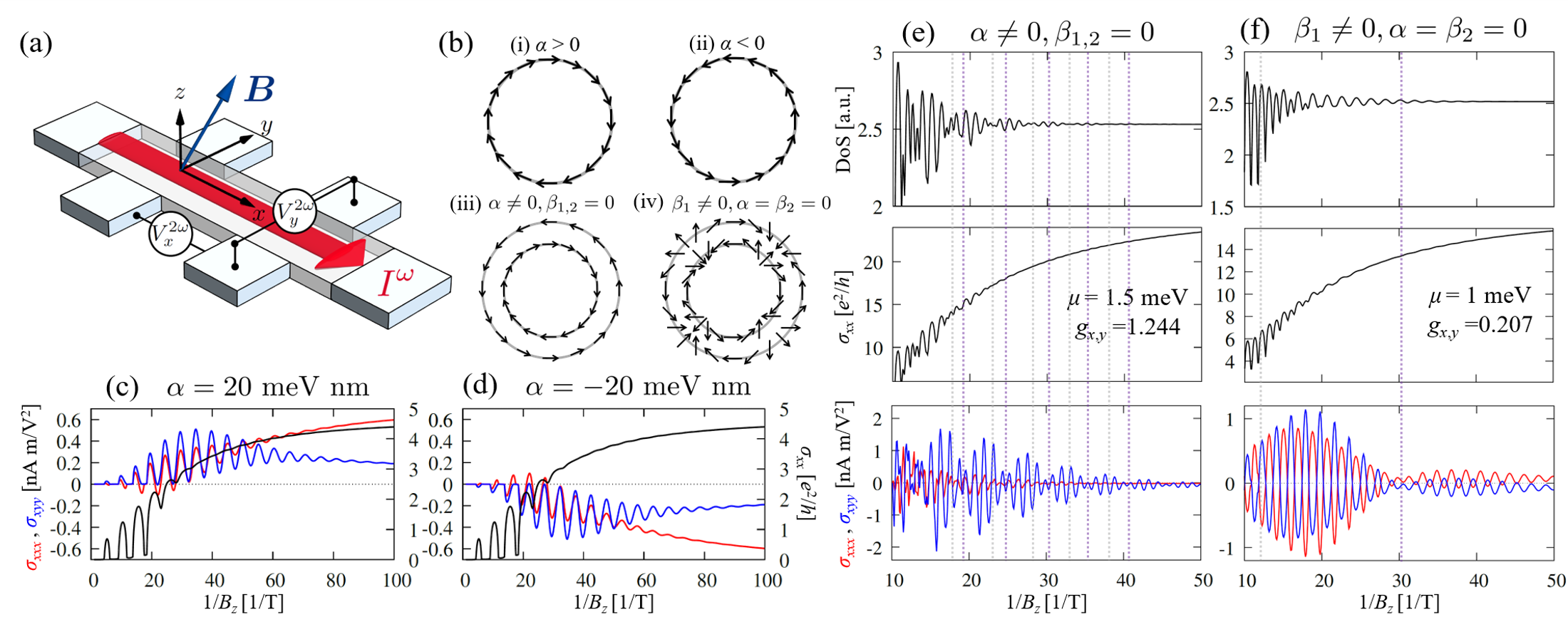}
\vspace*{-7mm}
\caption{
\textbf{Phenomenology of NSdH effect.} 
(a) Schematic of our setup. We assume a lock-in measurement in which a low-frequency ac current $I^\omega$ (corresponding to an ac electric field $E(\omega)$ in our theoretical framework) is injected, and the $2\omega$ components of the longitudinal and Hall voltages are read out. The relation between the measured second-harmonic voltages (nonlinear resistivities) and the nonlinear conductivities used below is discussed in the SM~\cite{com1}.
(b) Schematic illustrations of the in-plane spin textures on the Fermi surface at $B=0$. They correspond to (i,ii) single-Fermi-surface cases with pure $\alpha$ SOI for (i) $\alpha>0$ and (ii) $\alpha<0$, and (iii,iv) two-Fermi-surface cases for (iii) pure $\alpha$ and (iv) pure $\beta_1$, respectively.
(c,d) Linear (black lines) and nonlinear SdH oscillations (red: $\sigma_{xxx}$, blue: $\sigma_{xyy}$) for the model exhibiting a single Fermi surface with (c) $\alpha = 20$~meV~nm and (d) $\alpha = -20$~meV~nm. We set $m^* = \infty$, $\lambda = 20$~nm, $\beta_{1,2}= 0$~meV~nm$^3$, $g_{x,y} = 1.244$, $g_z=10$, $(B_x, B_y) = (0~{\rm T}, \, 0.1~{\rm T})$, and $\mu = 0.6$~meV in both cases. We set the impurity-scattering parameter to $n_{\rm i} u^2 = 100$~meV$^2$~nm$^2$.
(e,f) Density of states (DoS), and linear and nonlinear SdH oscillations for 2DHGs. We consider $\hbar^2 / (2m^*) = 620$~meV~nm$^2$, $\lambda = 0$, and $g_z = 10$ with (e) $\alpha = 1.5$~meV~nm, $\beta_{1,2} = 0$, $g_{x,y} = 1.244$, $\mu=1.5$~meV and (f) $\beta_1 = 190$~meV~nm$^3$, $\alpha = \beta_{2} = 0$, $g_{x,y} = 0.207$, $\mu=1$~meV, inspired by the Ge~[110] and Ge~[100] cases, respectively~\cite{Scappucci2021,XGLL2021,DLBLK,NLKL2024}.
We employ an in-plane magnetic field of $(B_x, B_y) = (0~{\rm T}, \, 5~{\rm T})$, and set the impurity parameter to $n_{\rm i} u^2 = 60$~meV$^2$~nm$^2$. Purple and gray dotted lines indicate the positions of the beating nodes in $\sigma_{xii}$ and $\sigma_{xx}$, respectively. We use a sufficiently large number of LLs, $N = 400 \sim 500$, in all panels (c)–(f).}
\label{fig:1}
\end{figure*}
For a finite out-of-plane magnetic field, the static kinetic momenta satisfy $[\hat\pi_x,\hat\pi_y]=i\hbar qB_z$. Introducing $\eta_B\equiv {\rm sgn}(qB_z)$, we define the ladder operators $\hat a = \frac{\ell_{\rm B}}{\sqrt{2}\hbar} \left( \hat\pi_x+i\eta_B\hat\pi_y \right)$ and $\hat a^\dagger = \frac{\ell_{\rm B}}{\sqrt{2}\hbar} \left( \hat\pi_x-i\eta_B\hat\pi_y \right)$, where $\ell_{\rm B}\equiv\sqrt{\hbar/|qB_z|}$.

\begin{figure*}[t]
\hspace*{-2mm}
\includegraphics[width=180mm]{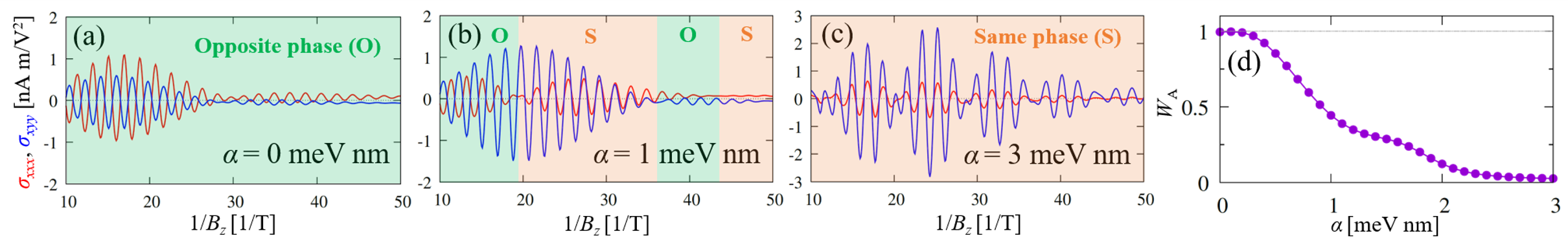}
\vspace*{-4mm}
\caption{
\textbf{Spin texture dependence of NSdH effect.} 
(a-c) Linear Rashba SOI dependence of the linear and nonlinear SdH oscillations for Ge [100] parameters  (i.e. $\alpha = 0$~meV~nm, $\beta_1 = 190$~meV~nm$^3$, $\beta_2 = 23.75$~meV~nm$^3$, and $g_{x,y} = 0.207$~\cite{Scappucci2021,XGLL2021,DLBLK,NLKL2024}) at a common chemical potential $\mu = 1$~meV, with (a) $\alpha = 0$, (b) $\alpha = 1$ and (c) $\alpha = 3$~meV~nm.
(d) $\alpha$ dependence of the effective phase-relation measure $W_{\rm A} = (1-\rho)/2$ between $\sigma_{xxx}$ and $\sigma_{xyy}$ defined by the correlation $\rho = \int \sigma_{xxx} (X) \sigma_{xyy} (X) dX/\sqrt{\int \sigma_{xxx}(X)^2 dX \int \sigma_{xyy}(X)^2 dX}$ for the Ge [100] parameters. Here $X\equiv1/B_z$, and all integrals are evaluated over the plotted interval $10\le X\le50~{\rm T}^{-1}$. For all panels in this figure, we assume a purely imaginary constant self-energy $\gamma=0.06~{\rm meV}$~\cite{Zhang2024,CSES2025}, and the remaining common parameters are the same as those in Figs.~\ref{fig:1}(e) and (f).
}
\label{fig:2}
\end{figure*}

For the numerical calculation, we represent the Hamiltonian and all vertex operators directly in the truncated LL--spin basis $\{\ket{n,s}\}$, with $n=0,\ldots,N-1$ and $s=\uparrow,\downarrow$. Thus, all operators are represented by $2N\times2N$ matrices. We evaluate the retarded and advanced Green-function matrices, $\hat G^{\rR}(\mu) = [ \mu\hat I-\hat H(\hat{\bm\pi}) -\hat\Sigma^{\rR}(\mu) ]^{-1}$ and $\hat G^{\rA}(\mu) = [\hat G^{\rR}(\mu) ]^\dagger$ with $\hat \Sigma^{\rm R}$ the retarded self-energy which we discuss in detail later. Working at $T=0$ K, the second-order conductivity tensor in the dc limit, defined by $j_i^{(2)}=\sigma_{ijl}E_jE_l$, is then written compactly as
\begin{align}
\sigma_{ijl}
&\simeq
\frac{\hbar^2}{2\pi}
N_{\rm L}\,
{\rm Im}\,
{\rm Tr}
\bigg[
\hat\Gamma_i
(\hat G^{\rR})^2
\nonumber\\
&\times
\Big(
\hat\Gamma_j\hat G^{\rR}\hat\Gamma_l
+
\hat\Gamma_l\hat G^{\rR}\hat\Gamma_j
+
2\hat\Gamma_{jl}
\Big)
(\hat G^{\rR}-\hat G^{\rA})
\bigg],
\label{eq:NLDF}
\end{align}
where ${\rm Tr}$ denotes the trace over the LL and spin degrees of freedom, and $N_{\rm L} = (2\pi\ell_{\rm B}^2)^{-1}$ is the Landau degeneracy per unit area. Equation~\eqref{eq:NLDF} is the LL representation 
derived from the Keldysh formalism and is analogous to the momentum-space expressions~\cite{MP,MN,NLKL2024,Shibata2026,com1}.
The matrix form in Eq.~\eqref{eq:NLDF} retains the LL and spin mixing contained in both the Hamiltonian and the self-energy.
For comparison, we calculate the linear longitudinal conductivity using the corresponding Kubo formula~\cite{Kubo1957,LK1958,YF2024},
\begin{align}
\sigma_{ii}
=
\frac{\hbar N_{\rm L}}{2\pi}
{\rm Re}\,
{\rm Tr}
\left[
\hat\Gamma_i
\hat G^{\rR}
\hat\Gamma_i
(\hat G^{\rA}-\hat G^{\rR})
\right].
\label{eq:linear}
\end{align}
For the calculations shown in Fig.~\ref{fig:1}, we evaluate the impurity self-energy within the self-consistent Born approximation (SCBA) for short-range spin-independent impurities~\cite{com1,Gilbertson2009,Rendell2023}. The SCBA self-energy is evaluated as a full $2\times2$ matrix in spin space. Denoting by $[\hat G^{\rR}]_{nn}$ the $2\times2$ spin block of the Green function associated with the $n^{\rm th}$ LL, the SCBA equation is
\begin{align}
\widetilde{\bm\Sigma}_{\rm s}^{\rR}(\mu)
=
2n_{\rm i}u^2N_{\rm L}
\sum_{n=0}^{N-1}
[\hat G^{\rR}(\mu)]_{nn}.
\label{eq:SCBA}
\end{align}
Here, $n_{\rm i}$ and $u$ are the impurity concentration and the strength of the impurity potential, respectively. The spin-independent real part of $\widetilde{\bm\Sigma}_{\rm s}^{\rR}$ is absorbed into the chemical potential, while its full spin-dependent structure, including the spin-off-diagonal components, is retained. An additional constant broadening $-i\g_{\rm c}\bm 1_2$ is included to account phenomenologically for other sources of level broadening, taking $\g_{\rm c}=0.01$~meV. 
For Fig.~\ref{fig:2}, where we focus on the systematic dependence of the NSdH response on the SOI parameters, we instead employ the purely imaginary constant self-energy $\hat\Sigma^{\rR}=-i\g\hat I_{2N}$.

{\it NSdH effect as a probe of spin textures}.---
To clarify the analytical connection to the spin texture, we first consider the momentum-space formulation without orbital Landau quantization. 
We consider the generic model $H_\k = h_\k + \bar \S_\k \cdot {\bm \s}$, where $h_\k$ is even in $\k$ and $\bar \S_\k = \S_\k + \bm \Delta$, with $\S_\k$ odd in $\k$ and $\bm \Delta$ the constant Zeeman term defined below; for strictly odd $\bar \S_\k$, the sum in Eq.~\eqref{eq:nlin} vanishes.  
For an in-plane field, treating $\bar \S_\k \cdot {\bm \s}$ perturbatively gives
\begin{align}
\sigma_{ijl} = \frac{2e^3}{\pi \hbar V} \sum_{\bm{k}} \mathcal{K}_{\bm{k}} \bar{\bm{S}}_{\bm{k}} \cdot \left[ \mathcal{V}_{(j} \partial_{l)} \partial_{i}
\bar{\bm{S}}_{\bm{k}} \right]  ,
 \label{eq:nlin}
\end{align} 
where $\d_i \equiv \d/\d k_i$, ${\cal V}_i = \d_i h_\k$, and ${\cal K}_\k = {\rm Im} [(G_\k^\rR)^3 G_\k^\rA]$, with $G_\k^{\rR (\rA)}$ the retarded (advanced) Green function of $h_\k$ and $V$ the system area (see SM~\cite{com1}). 
The parentheses denote symmetrization. Equation~\eqref{eq:nlin} shows that the leading nonlinear response probes the spin-texture curvature, whereas the exact finite-$B_z$ relations below follow independently from the LL formulation.

As a concrete example, we consider the NSdH in an effective model of a two-dimensional hole gas (2DHG) with linear and cubic Rashba SOI; the cubic term containing both isotropic and anisotropic interactions. Hereafter, we take $q= +e$ to simulate 2DHGs. The system is subject to a magnetic field with out-of-plane and in-plane components
to generate both Landau quantization and asymmetry in the electron spectrum. The corresponding Hamiltonian is
\begin{align}
\hat H &= \frac{\hat{\bm \Pi}^2}{2m^*} 
+ \frac{1}{2} \{ \tilde \a, ({\bm \sigma}\times \hat{\bm \Pi})\cdot \hat{z} \} + i \tilde \b_1 (\hat \Pi_-^3 \s_+ - \hat \Pi_+^3  \s_- ) 
\nonumber \\
&+ i \tilde \b_2 (\hat \Pi_+ \hat \Pi_- \hat \Pi_+ \s_+ - \hat \Pi_- \hat \Pi_+ \hat \Pi_- \s_- ) + {\bm \Delta} \cdot {\bm \s} , 
\label{eq:Ge_model}
\end{align} 
where $m^*$ is an effective mass, $\tilde \a \equiv (\a / \hbar) (1 + \lambda^2 \hat{\bm \Pi}^2 / \hbar^2)$~\cite{Fu2009}, and $\tilde \beta_{1,2} \equiv \beta_{1,2} / \hbar^3$ specify the strengths of the linear and cubic Rashba SOI, respectively; $\{\hat P, \hat Q\}$ the anticommutator of the operators $\hat P$ and $\hat Q$, $\hat \Pi_\pm \equiv \hat \Pi_x \pm i\hat \Pi_y$, ${\bm \s} = (\s_x, \s_y, \s_z)$ is the vector of Pauli matrices, $\sigma_{\pm}= (\sigma_x \pm i \sigma_y)/2$, and ${\bm \Delta} = (\mu_{\rm B}/2) (g_{x} B_x, g_{y} B_y, g_z  B_z)$ with $\mu_{\rm B}$ and $g_i$ denoting the Bohr magneton and effective $g$-factors, respectively. Here, Eq.~\eqref{eq:Ge_model} is a low-energy effective model for the lowest relevant subband, valid within appropriate ranges of $\mu$ and $\bm B$. The corresponding wave-vector representation is obtained by setting $h_{\bm k}=(\hbar k)^2/(2m^*)$ and $\overline{\bm S}_{\bm k}=\bm S_{\bm k}+\bm\Delta$, where $\bm S_{\bm k}=\bm S_{\alpha,\bm k}+\bm S_{\beta_1,\bm k}+\bm S_{\beta_2,\bm k}$ with
\begin{align}
\bm S_{\alpha,\bm k}&=\alpha k(1+\lambda^2k^2)(\sin\phi_{\bm k},-\cos\phi_{\bm k},0),\\
\bm S_{\beta_1,\bm k}&=\beta_1k^3(\sin3\phi_{\bm k},-\cos3\phi_{\bm k},0),\\
\bm S_{\beta_2,\bm k}&=-\beta_2k^3(\sin\phi_{\bm k},\cos\phi_{\bm k},0),
\end{align}
and $\tan\phi_{\bm k}=k_y/k_x$~\cite{GS1992,SL2005,SLBL2009,NKK2012,Moriya2014}. The three SOI terms generate distinct angular spin textures, with winding numbers $w_\alpha=+1$, $w_{\beta_1}=+3$, and $w_{\beta_2}=-1$, respectively (see SM~\cite{com1} for the further discussion). As seen in Eq.~\eqref{eq:nlin}, the curvature of the spin texture is relevant in the presence of the cubic terms. The nonlinear conductivity changes sign when the spin texture is inverted (i.e. $\S_\k \to -\S_\k$), as in previous studies~\cite{Ideue2017,Li2021,DLBLK,NLKL2024}.

First, we discuss the SdH and NSdH oscillations calculated using Eqs.~\eqref{eq:NLDF} and \eqref{eq:linear} for a single Fermi-surface with spin-momentum locking ($\beta_1=\beta_2=0$)~\cite{Fu2009}. Figs.~\ref{fig:1}(c) and (d) show the $B_z$ dependence of the linear longitudinal conductivity $\sigma_{xx}$ and the second-order nonlinear conductivity $\sigma_{xii}$ for two cases characterized by opposite signs of the linear Rashba parameter. 
Reversing $\alpha$ reverses the spin texture [Figs.~\ref{fig:1}(b)(i) and (ii)] and produces a simultaneous phase inversion of $\sigma_{xxx}$ and $\sigma_{xyy}$; the relation $\sigma_{xii}(\alpha)=-\sigma_{xii}(-\alpha)$ is proven in the SM~\cite{com1}. In contrast, linear SdH oscillations in $\s_{xx}$ are insensitive to the spin-texture on the Fermi-surface.

Next, we consider simplified effective models inspired by Ge-based planar heterostructures~\cite{Scappucci2021,DLBLK,NLKL2024,XGLL2021}. Figures~\ref{fig:1}(e) and (f) show the contrast between quantum oscillations for the pure-$\alpha$ and pure-$\beta_1$ models, respectively. Their spin textures have winding numbers $w_\alpha=+1$ and $w_{\beta_1}=+3$ [Figs.~\ref{fig:1}(b)(iii) and (iv)], providing a simple representative comparison between distinct SOI winding structures. In both models, when $B=0$, the spin textures of the inner and outer Fermi surfaces are oppositely oriented, reflecting the opposite spin expectation values of the two spin-split eigenbranches.

This example demonstrates the key features that make the NSdH effect a highly sensitive probe of spin textures. First, while both the linear SdH ($\sigma_{xx}$) and nonlinear NSdH ($\sigma_{xii}$) effects exhibit beating, their node positions alternate, which underscores how the two effects are influenced by spin-split bands. Linear SdH oscillations closely follow the total density of states, and their beating results from the interference of oscillatory contributions from the two Fermi surfaces with similar phases. In contrast, as discussed above, the NSdH effect is directly sensitive to the spin texture. Consistent with the branch-resolved analysis above (SM~\cite{com1}), the two spin-split branches carry opposite leading NSdH weights in the weak-$B_z$ intraband regime, so their interference shifts the nonlinear beating relative to the approximately branch-even linear SdH signal.
 
Furthermore, NSdH is not only sensitive to the presence and direction of a Fermi surface spin texture, but also provides a tool to distinguish SOIs with different symmetries. In particular, using the different components of the conductivity tensor, we can distinguish the linear and cubic Rashba SOI found in our Ge 2DHG example. For purely linear SOI ($\alpha$) the longitudinal $\sigma_{xxx}$ and Hall $\sigma_{xyy}$ nonlinear conductivities are in-phase, while for purely cubic SOI ($\beta_1$), they are perfectly out-of-phase [Figs.~\ref{fig:1}(e) and (f)]. This difference reflects the distinct symmetry structures of the SOI terms. In fact, for the pure-$\beta_1$ model, the LL selection rules associated with the cubic harmonicity enforce the exact relation (see SM~\cite{com1}):
\begin{equation}
    \sigma_{xxx} = - \sigma_{xyy}.
    \label{eq:relation}
\end{equation}
This relation provides a characteristic signature of the cubic $\beta_1$ SOI through the \lq\lq antisymmetric component'', $\sigma_{xxx}-\sigma_{xyy}$, of the conductivity tensor and evinces the utility of NSdH to characterize SOI.

By tuning the linear Rashba strength ($\alpha$) via gate voltages, we can directly track the evolution of the phase relation between $\sigma_{xxx}$ and $\sigma_{xyy}$, thereby probing the competition between linear and cubic SOI (Fig.~\ref{fig:2}). For representative Ge parameters, even after accounting for Joule-heating-induced thermal damping, we estimate equivalent second-harmonic currents of several tens of femtoamperes at experimentally accessible drive amplitudes, placing the predicted signal in a challenging but potentially accessible range for sensitive low-temperature measurements. The relation to the experimentally measured nonlinear resistivity is discussed in the SM~\cite{com1,Drichko2009,Price1982,Leturcq2003,Ando1982,Hatke2009,Levitin2022,Myronov2023,Kuntsevich2015,Tupikov2015,DelVecchio2026,Coppini2026}.

{\it Summary.}---We have established NSdH as a phase-sensitive probe of SOI that distinguishes, for example, linear and cubic Rashba couplings through their interference patterns and tensor symmetries. Although demonstrated here for Rashba-type Ge 2DHGs, the framework is applicable more broadly to spin-textured Fermi surfaces and provides a route to characterizing SOI in spintronic and quantum-information devices.

{\it Note added.}---Very recently, quantum oscillations in nonlinear electrical transport were experimentally observed in $\alpha$-Sn~\cite{Kalappattil2026}. This demonstrates the experimental accessibility of nonlinear quantum-oscillation measurements and provides additional motivation for the phase-resolved NSdH spectroscopy considered here.

{\it Acknowledgments.}---K.N. thanks H. Isobe, A. Yamada, and T. Yamaguchi for the helpful discussions. K.N. is supported by JSPS KAKENHI Grant Numbers JP21K13875 and JP26K07025. This work was supported as a part of NCCR SPIN, a National Centre of Competence in Research, funded by the Swiss National Science Foundation (grant number 225153). This publication is based upon work supported by King Fahd University of Petroleum \& Minerals. The author at KFUPM acknowledges the Deanship of Research and the Center for Advanced Quantum Computing for the support received under Grant no. CUP25102 and no. INQC2500, respectively.

{\it Data availability.}---The codes and data that support the findings of this article are openly available~\cite{NSdH_dataset}.

\clearpage

\setcounter{equation}{0}
\setcounter{figure}{0}
\setcounter{table}{0}
\setcounter{section}{0}
\setcounter{subsection}{0}
\setcounter{secnumdepth}{2}

\renewcommand{\theequation}{S\arabic{equation}}
\renewcommand{\thefigure}{S\arabic{figure}}
\renewcommand{\thesection}{\Roman{section}}

\allowdisplaybreaks[4]

\onecolumngrid

\begin{center}
\textbf{\large Supplemental Material: Probing Fermi-surface spin-textures via the nonlinear Shubnikov-de Haas effect}    
\end{center}

\section{Formulation of the nonlinear conductivity in the presence of Landau levels}

The statistical average of the electric current can be calculated within the Keldysh formalism:
\begin{align}
\langle \hat j_i \rangle = -\frac{i}{2} \int \frac{d\ve}{2\pi} {\rm tr} [ \hat j_i \hGK (\ve) ] . 
\end{align} 
The Hamiltonian and the current operator are expanded in powers of $\bm A'' (t)$ as
\begin{align}
\hat H(\hat{\bm \Pi}) &= \hat H(\hat {\bm \pi}) + \hG_i A''_i (t) + \hG_{ij} A''_i (t) A''_j (t) + \hG_{ijl} A''_i (t) A''_j(t) A''_l(t) + \cdots ,
\\
\hat{j}_i &= \hL_i + \hL_{ij} A''_j (t) + \hL_{ijl} A''_j (t) A''_l (t) + \cdots ,
\end{align}
where $\hG_i, \, \hG_{ij}, \hG_{ijl}, \cdots$ and $\hL_i, \, \hL_{ij}, \, \hL_{ijl}, \cdots$ are the expansion coefficients and the Einstein summation convention is employed for the spatial indices $i,j,l$.
$\hGK (\ve)$ can be expanded in terms of the coupling term $\hat H(\hat{\bm \Pi}) - \hat H(\hat{\bm \pi}) = \int \frac{d\w}{2\pi} \hX_i (-\w) F_i (\w)$ as 
\begin{align}
\hGK (\ve) &= \hat G^{\rm K(0)} (\ve) + \int \frac{d\w}{2\pi}  \hat G^{\rm K(1)} (\ve;\w) + \iint \frac{d\w}{2\pi} \frac{d\w'}{2\pi}  \hat G^{\rm K(2)} (\ve; \w, \w') ,
\\
\hat G^{\rm K(0)} (\ve)
&= (1 - 2f(\ve) ) ( \hat G^\rR(\ve) - \hat G^\rA (\ve) ) ,
\label{eq:K0a}
\\
\hat G^{\rm K(1)} (\ve;\w)
&= - 2 \Bigl[ \left( f_+ - f_- \right) \hGR_+ \hat{X}_i (-\w) \hGA_- 
+  f_- \hGR_+ \hat{X}_i (-\w) \hGR_- 
-  f_+ \hGA_+ \hat{X}_i (-\w) \hGA_- \Bigr] F_i (\w),
\label{eq:K1a}
\\
\hat G^{\rm K(2)} (\ve; \w, \w')
&=-\Bigl[
( f_{++} - f_{-+} ) \hat{G}_{++}^\rR \hat{X}_i (-\w) \hGA_{-+} \hat{X}_j (-\w') \hGA_{--} 
+ ( f_{-+} - f_{--} ) \hGR_{++} \hat{X}_i (-\w) \hGR_{-+} \hat{X}_j (-\w') \hGA_{--}
\nonumber \\
&\quad \quad
+ f_{++} \left\{ \hGA_{++} \hat{X}_i (-\w) \hGA_{-+} \hat{X}_j (-\w') \hGA_{--} \right\} 
- f_{--} \left\{ \hGR_{++} \hat{X}_i (-\w) \hGR_{-+} \hat{X}_j (-\w') \hGR_{--} \right\} 
\Bigr] 
F_i (\w) F_j (\w')
\nonumber \\
&\quad + (i,\w \leftrightarrow j,\w') .
\label{eq:K2a}
\end{align}
Here we define $F_i (\w) = A''_i (\w)$ and $\hX_i (-\w) = \hG_i (-\w) + \int \frac{d\w'}{2\pi} \, \hG_{ij} (-\w - \w') A''_j (\w') + \int \frac{d\w'}{2\pi} \int \frac{d\w''}{2\pi} \,\hG_{ijl} (-\w - \w' - \w'') A''_j (\w') A''_l (\w'') + \cdots$, and employ the notations $\ve_\pm = \ve \pm \hbar \w/2$, $\ve_{\pm\pm} = \ve \pm \hbar \w/2 \pm \hbar \w'/2$, $x_\pm \equiv x (\ve_\pm)$, and $x_{\pm \pm} \equiv x (\ve_{\pm\pm})$. Keeping terms up to second order in the time-dependent vector potential $\A''(t)$ and taking the Fourier components at $\w_1$ and $\w_2$, we obtain the second-order ac current $\langle \hat j_i^{(2)} \rangle (\w_1,\w_2)$ as
\begin{align}
\langle \hat j_i^{(2)} \rangle (\w_1,\w_2) &= 
\frac{-i}{2\w_1 \w_2} E_j (\w_1) E_l (\w_2)  \int \frac{d\ve}{2\pi}
\nonumber \\ 
&\quad \times {\rm tr} \Bigl[ f(\ve) \Bigl\{
2 \hG_{ijl} \GRmGA 
+ 
\hL_{ij} \hat{G}^\rR (\ve + \hbar \w_2) \hG_l \GRmGA + 
\hL_{ij} \GRmGA \hG_l \hat{G}^\rA (\ve - \hbar \w_2) \nonumber \\
&\quad \quad \quad \quad \quad \quad \quad \quad \quad \quad \quad \quad \quad \  
+ 
\hL_{il} \hat{G}^\rR (\ve + \hbar \w_1) \hG_j \GRmGA + \hL_{il} \GRmGA \hG_j \hat{G}^\rA (\ve - \hbar \w_1)  \nonumber \\
&\quad \quad \quad \quad \quad \quad \quad \quad \quad \quad \quad \quad \quad \ 
+ 
2\hL_{i} \hat{G}^\rR (\ve + \hbar \Omega) \hG_{jl} \GRmGA + 2\hL_{i} \GRmGA \hG_{jl} \hat{G}^\rA (\ve - \hbar \Omega) \nonumber \\
&\quad \quad + \hL_i \hat{G}^\rR (\ve + \hbar \Omega) \hG_j \hat{G}^\rR (\ve + \hbar \w_2) \hG_l \GRmGA + \hL_i \hat{G}^\rR (\ve + \hbar \Omega) \hG_l \hat{G}^\rR (\ve + \hbar \w_1) \hG_j \GRmGA \nonumber \\
&\quad \quad + \hL_i \hat{G}^\rR (\ve + \hbar \w_1) \hG_j \GRmGA \hG_l \hat{G}^\rA (\ve - \hbar \w_2) + \hL_i \hat{G}^\rR (\ve + \hbar \w_2) \hG_l \GRmGA \hG_j \hat{G}^\rA (\ve - \hbar \w_1) \nonumber \\
&\quad \quad + \hL_i \GRmGA \hG_j \hat{G}^\rA (\ve - \hbar \w_1) \hG_l \hat{G}^\rA (\ve - \hbar \Omega) + \hL_i  \GRmGA \hG_l \hat{G}^\rA (\ve + \hbar \w_1) \hG_j  \hat{G}^\rA (\ve - \hbar \Omega) \Bigr\} \Bigr] \nonumber \\
&\equiv \sigma_{ijl} (\w_1,\w_2) E_j (\w_1) E_l (\w_2),
\label{eq:ja}
\end{align} 
where $\Omega = \w_1 + \w_2$ and we omit the $\ve$ dependence in $\GRmGA$ for simplicity. 
When we take the dc limit, some $1/\w$-type terms appear; however, these terms can be disregarded because of causality (analytic continuation)~\cite{Kubo1957,Shibata2026}. Hence, the expression in the dc limit is  
\begin{align}
\langle \hat{j}_i^{(2)} \rangle &= \sigma_{i j l} E_j E_l  = ( \sigma_{ijl}^{\rm DF} + \sigma_{ijl}^{\rm F} ) E_j E_l , \label{eq:DC}  \\
\sigma_{ijl}^{\rm DF}
&=
 \hbar^2 \int \frac{d\ve}{2\pi} \left(-\frac{\partial f}{\partial \ve}\right) 
{\rm Im} \left[
{\rm tr} \Biggl\{ \hL_i  
 \frac{\partial \hat{G}^\rR }{\partial \ve} \left( \hG_j \hat{G}^\rR \hG_l + \hG_{jl} \right) ( \hat{G}^\rR - \hat{G}^\rA ) \Biggr\} \right] 
 + (j \leftrightarrow l), 
 \label{eq:NLSur}
 \\
 \sigma_{ijl}^{\rm F} &=  -\hbar^2 \int \frac{d\ve}{2\pi} f(\ve) {\rm Im} \left[
 {\rm tr} 
 \left\{ 
 \hL_i \frac{\partial \hat{G}^\rR}{\partial \ve} \left( \hG_j \hat{G}^\rR \hG_l + \hG_{jl} \right) \frac{\partial \hat{G}^\rR}{\partial \ve} \right\} \right] + (j \leftrightarrow l) .
\label{eq:NLC}
\end{align}
This is analogous to previously derived formulas written in the wavenumber representation~\cite{Shibata2026,MP,MN,NLKL2024}. We here focus on the $T=0$ case, and a detailed analysis of the extraction of a Dingle temperature, as in the Lifshitz--Kosevich treatment~\cite{LK1956,LK1958}, lies beyond the scope of the present Letter. Although it is difficult to explicitly discuss the effects of quantum geometry here because we consider Landau levels, the momentum-space degrees of freedom are replaced by those of the cyclotron motion and guiding center, so the calculated results still incorporate quantum-geometric effects. In addition, the Fermi-sea term $\sigma^{\rm F}_{ijl}$ vanishes with the damping in the clean dc limit and is therefore neglected here~\cite{MP,MN}. Thus, at $T=0$ and retaining the Fermi-surface contribution, we arrive at the expression given in Eq.~(2) of the main text.

\section{Convention for the hole gas and the sign of $B_z$.}

In the numerical calculation, we use a common Landau-level convention in
which the orbital magnetic-field scale is specified by
$B_z^{\rm code}=|B_z|>0$.  Physically, the 2DEG and 2DHG calculations
correspond to
\begin{equation}
\begin{array}{c|ccc}
 & q & B_z^{\rm phys} & \Delta_z \\ \hline
{\rm 2DEG} & -e & +|B_z| &
+\dfrac{\mu_B g_z}{2}|B_z|
\\[2mm]
{\rm 2DHG} & +e & -|B_z| &
-\dfrac{\mu_B g_z}{2}|B_z|
\end{array}.
\end{equation}
Thus, in both cases,
\begin{equation}
qB_z^{\rm phys}=-e|B_z|,
\end{equation}
and the same Landau-level ladder convention can be used.  Accordingly,
the orbital quantities such as
\begin{equation}
\ell_B=\sqrt{\frac{\hbar}{e|B_z|}},
\quad
\omega_c=\frac{e|B_z|}{m^*},
\end{equation}
are identical in the two implementations, while the sign of the
out-of-plane Zeeman term is reversed:
\begin{equation}
\Delta_z^{\rm 2DEG}
=
+\frac{\mu_B g_z}{2}B_z^{\rm code},
\quad
\Delta_z^{\rm 2DHG}
=
-\frac{\mu_B g_z}{2}B_z^{\rm code}.
\end{equation}
For the field configuration considered here,
$\bm B=(0,B_y,B_z)$, the mirror operation
$M_y:(x,y,z)\rightarrow(x,-y,z)$ transforms the axial magnetic field as
\begin{equation}
(B_x,B_y,B_z)
\longrightarrow
(-B_x,B_y,-B_z).
\end{equation}
Hence, for $B_x=0$, it relates the systems at $+B_z$ and $-B_z$ while
leaving $B_y$ unchanged.  Since $j_x$, $E_x^2$, and $E_y^2$ are even
under this mirror operation, the longitudinal response coefficients
considered in this work satisfy
\begin{equation}
\sigma_{xx}(B_y,B_z)
=
\sigma_{xx}(B_y,-B_z),
\end{equation}
\begin{equation}
\sigma_{xxx}(B_y,B_z)
=
\sigma_{xxx}(B_y,-B_z),
\quad
\sigma_{xyy}(B_y,B_z)
=
\sigma_{xyy}(B_y,-B_z).
\end{equation}
Therefore, the sign of the physical perpendicular magnetic field does
not affect these conductivity components, provided that the sign of
the associated Zeeman term is transformed consistently.  In the
present code this is implemented by using the same positive
$B_z^{\rm code}=|B_z|$ for the orbital Landau-level sector and changing
only the sign of $\Delta_z$ between the 2DEG and 2DHG calculations.

\section{On the comparison between SCBA and imaginary constant self energy}

\begin{figure}[t]
\includegraphics[width=180mm]{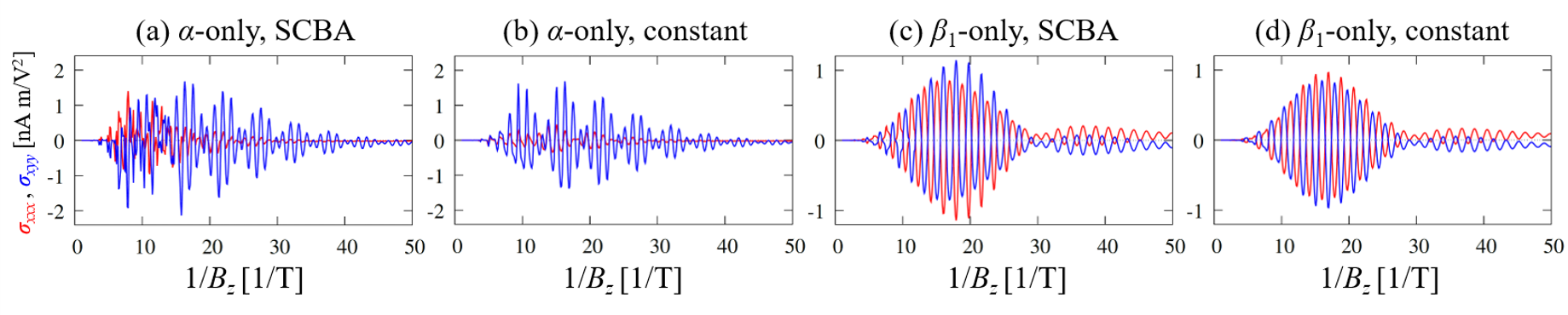}
\vspace*{-8mm}
\caption{The nonlinear SdH oscillations in (a,b) $\alpha$-only system at $\mu=1.5$~meV and (c,d) $\beta_1$-only system at $\mu=1$~meV in (a,c) self-consistent Born approximation (SCBA) with $n_{\rm i} u^2 = 60~{\rm meV^2~nm^2}$ and (b,d) a purely imaginary constant self-energy $\Sigma^{\rm R}=-i\gamma$ with $\gamma=0.07$~meV}
\label{fig:S1}
\end{figure}

Figure~\ref{fig:S1} shows the oscillations in $\sigma_{xii}$, comparing calculations performed using the self-consistent Born approximation (SCBA) and a purely imaginary constant self-energy. For the SCBA calculations, the self-energy iteration is initialized with the spin-independent purely imaginary seed $-i(0.01~{\rm meV})\hat I_2$ and performed with a linear mixing parameter of $0.3$ until the convergence criterion $\|\widetilde{\Sigma}^{R}_{s,n+1}-\widetilde{\Sigma}^{R}_{s,n}\|_{\infty}<10^{-10}\max\!\left[1,\|\widetilde{\Sigma}^{R}_{s,n+1}\|_{\infty}\right]$ is satisfied, with a maximum of 1000 iterations at each $B_z$. All magnetic-field sweeps are performed on a uniform inverse-field grid with $\Delta(1/B_z)=0.1~{\rm T}^{-1}$. The SCBA results become more complex in the high-field regime, particularly in the $\alpha$-only case, where additional fine
structures appear. Nevertheless, the characteristic phase relations and beating structures are preserved, and the two treatments show no qualitative difference in the central features discussed in the main text.

Note that we neglect impurity vertex corrections in the numerical calculations presented here; such corrections may quantitatively renormalize the nonlinear conductivity. For the spin-independent white-noise disorder considered here, however, the corresponding ladder corrections preserve the Landau-level-index structure modulo three and hence the selection rules underlying the exact pure-$\beta_1$ relation $\sigma_{xxx}=-\sigma_{xyy}$. A detailed proof is given in Sec.~\ref{sec:vcproof}.

\section{Arguments from the analytical calculations of the nonlinear conductivity}

\subsection{Momentum space analysis under the in-plane magnetic field}

Here, we reexamine the calculation of the second-order nonlinear conductivity in the presence of an in-plane magnetic field~\cite{Ideue2017,Li2021,DLBLK,NLKL2024} to discuss the relation between the spin texture on the band structure and the nonlinear response. We start from the $T=0$ expression: 
\begin{align}
&\sigma_{ijl}
\simeq
-\frac{e^3}{2 \pi \hbar V} 
{\rm Im} \sum_\k  
{\rm tr} \left\{ \cV_i  
 ({\cal G}^\rR)^2 \left( \cV_j {\cal G}^\rR \cV_l + \cV_l {\cal G}^\rR \cV_j + \cV_{jl} \right) {\cal G}^\rA \right\}  , 
\end{align}
where ${\cal G}^\rR \equiv (\mu - H_\k + i\g)^{-1} = ({\cal G}^\rA)^\dagger$ is retarded Green function with a constant purely imaginary self-energy $\hat{\Sigma}^{\rm R} = -i\hbar/(2\tau) = -i\gamma$, $\cV_{i} \equiv \partial_{i} H_\k$, and $\cV_{ij} \equiv \partial_{j} \cV_{i}$~\cite{MP,MN,NLKL2024}. The trace runs over the spin space. To clarify the dependence on the spin texture, we split the Hamiltonian into two parts as $H_\k = h_\k + \bar \S_\k \cdot {\bm \s}$ and assume a large $\mu$ and/or $h_\k$ for the perturbative treatment with respect to
$\bar \S_\k \cdot {\bm \s}$. We also assume that $\S_\k$ should be an odd function of $k_i$. The perturbative expansion of the retarded/advanced Green function is expressed as
\begin{align}
{\cal G}^{\rm R(A)} = G^{\rm R(A)} + G^{\rm R(A)} ( \bar \S_\k \cdot {\bm \sigma} ) G^{\rm R(A)} + G^{\rm R(A)} ( \bar \S_\k \cdot {\bm \sigma} ) G^{\rm R(A)} ( \bar \S_\k \cdot {\bm \sigma} ) G^{\rm R(A)} + \cdots,  
\end{align}
where $G^{\rR(\rA)} \equiv [\mu - h_\k \pm i\g ]^{-1}$. The first-order perturbation vanishes because of ${\rm tr} \, \sigma_i = 0$. The second-order contribution is crucial in our case. After  straightforward calculations, we find 
\begin{align}
\sigma_{ijl}
= 
\frac{e^3}{\pi\hbar V}
{\rm Im}
\sum_\k 
\Biggl[
(\d_l h_\k ) (\partial_j \partial_i \bar \S_\k) \cdot \bar \S_\k
+ (j \leftrightarrow l)
\Biggr]
(G^\rR)^3 G^\rA 
.
\label{eq:supplnlin}
\end{align} 
The second derivative of $\bar S_\k$ in Eq.~\eqref{eq:supplnlin} represents the curvature of the spin texture, which is relevant when the SOI term contains more than second-order of $\k$. This expression connects the
curvature of spin texture on the band structure and the second-order nonlinear conductivity. 

\subsection{Nonperturbative proof of $\sigma_{xii} (\alpha) = -\sigma_{xii} (-\alpha)$ in $\alpha$-only system}
\label{sec:alphaexact}

For the pure-$\alpha$ model with the cubic correction ($\lambda$ term), the sign reversal of the nonlinear response under inversion of the spin texture can also be established exactly without a perturbative expansion. Let $\hat{\mathcal P}_o$ denote orbital parity, $\hat{\mathcal P}_o\bm r\hat{\mathcal P}_o^{-1}=-\bm r$, acting trivially on the pseudospin. For a uniform perpendicular magnetic field, one may choose the symmetric gauge such that $\bm A'(-\bm r)=-\bm A'(\bm r)$, giving $\hat{\mathcal P}_o\bm\pi\hat{\mathcal P}_o^{-1}=-\bm\pi$. Including the spatially uniform time-dependent vector potential, this becomes $\hat{\mathcal P}_o\bm\Pi[\bm A'']\hat{\mathcal P}_o^{-1}=-\bm\Pi[-\bm A'']$. Since the kinetic and Zeeman terms are even under this transformation whereas the pure Rashba term is odd in $\bm\Pi$, one obtains
\begin{equation}
\hat{\mathcal P}_o\hat H_{\alpha}[\bm A'']\hat{\mathcal P}_o^{-1}
=
\hat H_{-\alpha}[-\bm A''].
\end{equation}
Consequently,
\begin{equation}
\hat{\mathcal P}_o\hat\Gamma_i(\alpha)\hat{\mathcal P}_o^{-1}
=-\hat\Gamma_i(-\alpha),\quad
\hat{\mathcal P}_o\hat\Gamma_{ij}(\alpha)\hat{\mathcal P}_o^{-1}
=+\hat\Gamma_{ij}(-\alpha),
\end{equation}
while $\hat{\mathcal P}_o\hat G_{\alpha}^{R,A}\hat{\mathcal P}_o^{-1}=\hat G_{-\alpha}^{R,A}$. Substitution into Eq.~(2) of the main text therefore gives the exact relation
\begin{equation}
\sigma^{(2)}_{ijl}(\alpha;B_y,B_z)
=
-\sigma^{(2)}_{ijl}(-\alpha;B_y,B_z),
\end{equation}
whereas the corresponding linear conductivity is even in $\alpha$. Thus, the simultaneous phase inversion of $\sigma_{xxx}$ and $\sigma_{xyy}$ in Figs.~1(c) and 1(d) follows directly from the reversal of the Rashba spin texture and remains valid in the Landau-quantized regime.

\subsection{Branch-resolved origin of the weak-field phase difference}
\label{sec:branch_phase}

The exact relation derived in Sec.~IV B concerns a global inversion of the spin texture in the pure-$\alpha$ model. We now show how the same sign sensitivity leads to oppositely signed contributions from the two spin-split Fermi surfaces in the weak-$B_z$ regime. This analysis is restricted to the intraband regime, in which the two Fermi contours can be treated as distinct semiclassical orbits and their interference gives rise to beating. We further assume that the spin splitting is small on the energy scale over which the smooth intraband response kernel varies, while the difference of the oscillation phases $2\pi(F_+-F_-)/B_z$ need not be small. The latter distinction allows two nearby Fermi surfaces to produce well-resolved beating even when their smooth response weights are nearly identical.

For the momentum-space Hamiltonian $\hat H_{\bm k}=h_{\bm k}+\overline{\bm S}_{\bm k}\cdot\boldsymbol{\sigma}$, with $\overline{\bm S}_{\bm k}=\bm S_{\bm k}+\boldsymbol{\Delta}$, we introduce $d_{\bm k}=|\overline{\bm S}_{\bm k}|$ and the band index $\eta=\pm$. The eigenenergies and Green functions can be written as
\begin{equation}
 \epsilon_{\eta\bm k}=h_{\bm k}+\eta d_{\bm k},\quad
 G^\rR_{\bm k}=\sum_{\eta=\pm}P_{\eta\bm k}g^\rR_{\eta\bm k},\quad
 P_{\eta\bm k}=\frac{1+\eta\overline{\bm S}_{\bm k}\cdot{\bm \sigma}}{2},
\end{equation}
where $g^\rR_{\eta\bm k}=(\mu-\epsilon_{\eta\bm k}+i\gamma)^{-1}$. Retaining only the intraband terms of the microscopic Keldysh expression gives~\cite{NLKL2024}
\begin{equation}
 \sigma^{\rm intra}_{ijl,\eta}
 ={\cal C}\sum_{\bm k}v_{\eta i}v_{\eta,jl}{\cal A}_{\eta\bm k},
 \qquad
 v_{\eta i}=\partial_i\epsilon_{\eta\bm k},\quad
 v_{\eta,jl}=\partial_j\partial_l\epsilon_{\eta\bm k},
 \label{eq:branch_intra}
\end{equation}
where ${\cal A}_{\eta\bm k}=-\mathrm{Im}\,g^\rR_{\eta\bm k}$ and ${\cal C}$ is a common prefactor. This expression is the intraband sector of the Keldysh formula and reduces to the usual Boltzmann expression in the clean limit~\cite{NLKL2024}. To leading order in the spin splitting of the smooth kernel, we write $V_i=\partial_i h_{\bm k}$, $V_{jl}=\partial_j\partial_l h_{\bm k}$, and ${\cal A}_{0\bm k}=-\mathrm{Im}(\mu-h_{\bm k}+i\gamma)^{-1}$. Equation~\eqref{eq:branch_intra} then gives
\begin{align}
 \sigma^{\rm intra}_{ijl,\eta}
 ={\cal C}\sum_{\bm k}V_iV_{jl}{\cal A}_{0\bm k}
 +\eta{\cal C}\sum_{\bm k}\Big[
 (\partial_i d_{\bm k})V_{jl}{\cal A}_{0\bm k}
 +V_i(\partial_j\partial_l d_{\bm k}){\cal A}_{0\bm k} 
 +V_iV_{jl}d_{\bm k}\partial_h{\cal A}_{0\bm k}
 \Big]+\cdots .
 \label{eq:branch_expand}
\end{align}
Since $h_{\bm k}$ is even in $\bm k$, the first term in Eq.~\eqref{eq:branch_expand} is odd and vanishes after momentum integration. Decomposing $d_{\bm k}=d^{\rm e}_{\bm k}+d^{\rm o}_{\bm k}$, where
\begin{equation}
 d^{\rm o}_{\bm k}
 =\frac{1}{2}\left(|\bm S_{\bm k}+\boldsymbol{\Delta}|-|\bm S_{\bm k}-\boldsymbol{\Delta}|\right),
 \qquad d^{\rm o}_{-\bm k}=-d^{\rm o}_{\bm k},
\end{equation}
parity similarly removes all terms containing $d^{\rm e}_{\bm k}$. Hence, to leading order,
\begin{equation}
 \sigma^{\rm intra}_{ijl,\eta}=\eta\,{\tilde \sigma}_{ijl}[d^{\rm o}]+\cdots ,
 \label{eq:branch_opposite}
\end{equation}
with the same functional ${\tilde \sigma}_{ijl}$ for the two branches. Thus, the two spin-split Fermi surfaces carry opposite leading nonlinear weights. Moreover, $\bm S_{\bm k}\rightarrow-\bm S_{\bm k}$ gives $d^{\rm o}_{\bm k}\rightarrow-d^{\rm o}_{\bm k}$, connecting Eq.~\eqref{eq:branch_opposite} directly to the spin-texture sign inversion discussed in Sec.~\ref{sec:alphaexact}. A weak perpendicular field quantizes the two Fermi contours separately. Applying the usual Poisson-resummation/Lifshitz--Kosevich argument to each branch~\cite{LK1958,Shoenberg1984}, Eq.~\eqref{eq:branch_opposite} therefore gives for the fundamental oscillation
\begin{equation}
 \delta\sigma^{(2)}_{ijl}
 \simeq
 b_+\cos\left(\frac{2\pi F_+}{B_z}+\phi_+\right)
 -
 b_-\cos\left(\frac{2\pi F_-}{B_z}+\phi_-\right),
 \qquad b_\pm>0.
 \label{eq:branch_qo}
\end{equation}
For the planar spin textures considered here, the Berry phases of the two branches differ by $2\pi$ times the winding number in the $B_z\rightarrow0$ limit and are therefore equivalent modulo $2\pi$; the Maslov contribution is also common. Hence $\phi_+\simeq\phi_-$ at weak $B_z$, and the opposite signs in Eq.~\eqref{eq:branch_qo} correspond to an approximately $\pi$ relative phase of the nonlinear oscillations. In contrast, the leading linear-conductivity weights are branch even. For comparable branch amplitudes, writing $\theta_\pm=2\pi F_\pm/B_z+\phi$ gives
\begin{equation}
 \delta\sigma^{(1)}\propto 2a\cos\frac{\theta_++\theta_-}{2}\cos\frac{\theta_+-\theta_-}{2},
 \quad
 \delta\sigma^{(2)}\propto -2b\sin\frac{\theta_++\theta_-}{2}\sin\frac{\theta_+-\theta_-}{2}.
\end{equation}
The linear and nonlinear beating nodes therefore alternate by half a beating period. If $F_+=F_-$, the leading branch-odd nonlinear contributions cancel instead of producing beating. Unequal branch amplitudes, finite-$B_z$ corrections, interband terms, and branch-dependent scattering can shift or broaden the nodes, which is why the relative $\pi$ phase is an asymptotic weak-$B_z$ result rather than an exact finite-field identity.

\subsection{Dependence on the SOI winding structure}

The three SOI terms in Eq.~(6) of the main text generate different angular spin textures. In the zero-field momentum representation, their winding numbers are $w_\alpha=+1$, $w_{\beta_1}=+3$, and $w_{\beta_2}=-1$. In particular, although both $\alpha$ and $\beta_2$ produce a single rotation of the spin texture around the Fermi surface in magnitude, their winding directions are opposite. Because the nonlinear conductivity depends on momentum derivatives of the spin texture, as explicitly seen in Eq.~\eqref{eq:supplnlin}, SOI terms with different winding structures generally produce different combinations of nonlinear tensor components. Thus, in principle, measurements of several nonlinear-conductivity components can provide information that distinguishes SOI contributions with different windings, rather than only their overall spin-splitting strengths.

In the representative Ge [100] parameter set considered in the main text, the isotropic and anisotropic cubic Rashba coefficients are $\beta_1=190~{\rm meV\,nm^3}$ and $\beta_2=23.75~{\rm meV\,nm^3}$, respectively, so that $|\beta_2/\beta_1|=0.125$. The $\beta_2$ term is therefore subdominant in the adopted parameterization, although it is retained in the realistic multi-SOI calculations of Fig.~2. For this reason, the main text uses the pure-$\alpha$ and pure-$\beta_1$ limits as the simplest representative comparison between winding-$1$ and winding-$3$ spin textures, while Fig.~2 explicitly includes the additional $\beta_2$ contribution when discussing realistic Ge parameters.

\subsection{On $\sigma_{xxx} = -\sigma_{xyy}$ for $\a = \b_2 =0$ case. }
\label{sec:pert}

In the main text, we show that the oscillations in $\sigma_{xxx}$ and $\sigma_{xyy}$ have mutually opposite phases for $\alpha=\beta_2=0$ and $\beta_1\neq0$, as expressed by the exact relation
\begin{equation}
    \sigma_{xxx}=-\sigma_{xyy}.
    \label{eq:sigmarelation}
\end{equation}
Here, we derive this relation directly in the Landau-level representation, without performing a perturbative expansion in $\beta_1$, the Zeeman field, or $B_z$. We decompose the Hamiltonian as $\hat H=\hat H_0+\hat H_{{\rm Z},\parallel}$, where $\hat H_0$ contains the kinetic term, the pure-$\beta_1$ SOI, and the out-of-plane Zeeman term $\Delta_z\sigma_z$, while $\hat H_{{\rm Z},\parallel}=\Delta_x\sigma_x+\Delta_y\sigma_y$ is the in-plane Zeeman interaction. Introducing the block basis $\mathcal{B}_{\ell}=(|\ell,\uparrow\rangle,|\ell+3,\downarrow\rangle)^{\mathsf T}$, the nontrivial $2\times2$ blocks of $\hat H_0$ are
\begin{equation}
    \hat H_{0}^{\ell,\ell}=
    \begin{pmatrix}
        \hbar\omega_{\rm c}\left(\ell+\frac{1}{2}\right)+\Delta_z & i\widetilde L_{\ell+1}\\
        -i\widetilde L_{\ell+1} & \hbar\omega_{\rm c}\left(\ell+\frac{7}{2}\right)-\Delta_z
    \end{pmatrix},
    \label{eq:H0beta1block}
\end{equation}
where $\widetilde L_{\ell}=R_{\beta_1}\sqrt{\ell(\ell+1)(\ell+2)}$ and $R_{\beta_1}=(\sqrt{2}/\ell_B)^3\beta_1$. The three lowest down-spin Landau states, for which the corresponding up-spin partner would have a negative Landau-level index, can be treated as one-dimensional boundary blocks; they obey the same block-index selection rules used below and do not modify the argument.

For any operator $\hat X$, we denote by $\hat X^{r,s}$ the block mapping $\mathcal{B}_s$ to $\mathcal{B}_r$. In this basis, the nonzero adjacent-block matrix elements of the first-order vertices are
\begin{align}
    \hat\Gamma_x^{\ell-1,\ell}
    =
    \begin{pmatrix}
        P\sqrt{\ell} & 0\\
        -i\widetilde L'_{\ell+1} & P\sqrt{\ell+3}
    \end{pmatrix},
    \quad
    \hat\Gamma_y^{\ell-1,\ell}
    =
    \begin{pmatrix}
        iP\sqrt{\ell} & 0\\
        \widetilde L'_{\ell+1} & iP\sqrt{\ell+3}
    \end{pmatrix},
    \label{eq:Gammaundirectedbeta1}
\end{align}
where
\begin{equation}
    P=\frac{e\hbar}{\sqrt{2}m^*\ell_B},
    \qquad
    \widetilde L'_{\ell}=3R'_{\beta_1}\sqrt{\ell(\ell+1)},
    \qquad
    R'_{\beta_1}=\frac{2e\beta_1}{\hbar\ell_B^2}.
\end{equation}
The relevant second-order vertices are
\begin{align}
    \hat\Gamma_{xx}^{\ell,\ell}
    =
    \hat\Gamma_{yy}^{\ell,\ell}
    =
    \frac{e^2}{2m^*}\hat I_2,
    \quad 
    \hat\Gamma_{xx}^{\ell-1,\ell+1}
    =
    \begin{pmatrix}
        0 & 0\\
        -3iR''_{\beta_1}\sqrt{\ell+2} & 0
    \end{pmatrix},
    \quad 
    \hat\Gamma_{yy}^{\ell-1,\ell+1}
    =
    \begin{pmatrix}
        0 & 0\\
        3iR''_{\beta_1}\sqrt{\ell+2} & 0
    \end{pmatrix},
    \label{eq:Gamma2beta1}
\end{align}
with
\begin{equation}
    R''_{\beta_1}=\frac{\sqrt{2}e^2\beta_1}{\hbar^2\ell_B}.
\end{equation}
The in-plane Zeeman term connects blocks whose indices differ by three:
\begin{equation}
    \hat H_{{\rm Z},\parallel}^{\ell-1,\ell+2}
    =
    \begin{pmatrix}
        0 & 0\\
        \Delta_x+i\Delta_y & 0
    \end{pmatrix}.
    \label{eq:HZparallelblock}
\end{equation}
The reverse blocks are fixed by Hermiticity, $\hat X^{s,r}=(\hat X^{r,s})^\dagger$, so that the complete operators $\hat H$, $\hat\Gamma_x$, $\hat\Gamma_y$, $\hat\Gamma_{xx}$, and $\hat\Gamma_{yy}$ are Hermitian.

Equation~\eqref{eq:H0beta1block} shows that $\hat H_0$ preserves the block index, whereas Eq.~\eqref{eq:HZparallelblock} changes it only by $\pm3$. The constant self-energy and the full $2\times2$ spin-space SCBA self-energy considered in this work obey the same modulo-three block-selection rule: their spin-diagonal components preserve the block index and their spin-off-diagonal components change it by $\pm3$. Consequently,
\begin{equation}
    \left(\hat G^{\rR,\rA}\right)^{r,s}=0
    \quad\text{unless}\quad
    r-s=3m,\quad m\in\mathbb Z,
    \label{eq:Gmod3selection}
\end{equation}
and the same relation holds for $\partial_\varepsilon\hat G^{\rR,\rA}$. By contrast, $\hat\Gamma_x$, $\hat\Gamma_y$, and the current vertex $\hat\Lambda_x=-\hat\Gamma_x$ change the block index by $\pm1$, while $\hat\Gamma_{xx}$ and $\hat\Gamma_{yy}$ change it by $0$ or $\pm2$.

For $\sigma_{xaa}$, with $a=x,y$, the $a$-dependent part of the Fermi-surface kernel in Eq.~\eqref{eq:NLSur} is $\hat{\mathcal K}_a=\hat\Gamma_a\hat G^\rR\hat\Gamma_a+\hat\Gamma_{aa}$. Since all Green functions preserve the block index modulo three and the outer current vertex $\hat\Lambda_x$ changes it by $\pm1$, the trace can close only through components of $\hat{\mathcal K}_a$ satisfying $r-s\equiv\pm2\pmod 3$. For the two-vertex term, this condition requires the two first-order vertices to change the block index in the same direction. From Eq.~\eqref{eq:Gammaundirectedbeta1}, the first-order vertices satisfy
\begin{equation}
    \hat\Gamma_y^{r,s}=-i(r-s)\hat\Gamma_x^{r,s},
    \quad r-s=\pm1.
    \label{eq:GammayGammaxphase}
\end{equation}
The two phase factors in every contributing path therefore multiply to $-1$, independently of the intermediate Green-function block, and hence
\begin{equation}
    \left(\hat\Gamma_y\hat G^\rR\hat\Gamma_y\right)^{r,s}
    =
    -\left(\hat\Gamma_x\hat G^\rR\hat\Gamma_x\right)^{r,s},
    \quad r-s\equiv\pm2\pmod 3.
    \label{eq:twovertexminus}
\end{equation}
For the contact term, Eq.~\eqref{eq:Gamma2beta1} gives
\begin{equation}
    \hat\Gamma_{yy}^{r,s}=-\hat\Gamma_{xx}^{r,s},
    \quad r-s=\pm2,
    \label{eq:contactminus}
\end{equation}
whereas the equal diagonal components $\hat\Gamma_{xx}^{r,r}=\hat\Gamma_{yy}^{r,r}$ cannot contribute because they do not satisfy the trace-closing selection rule. It follows that
\begin{equation}
    \hat{\mathcal K}_y^{r,s}=-\hat{\mathcal K}_x^{r,s}
    \label{eq:kernelminus}
\end{equation}
for every block sector contributing to the trace. All remaining factors in Eq.~\eqref{eq:NLSur} are identical for $\sigma_{xxx}$ and $\sigma_{xyy}$, and the same block-selection argument applies to the Fermi-sea term. We therefore obtain, without any perturbative treatment,
\begin{equation}
    \sigma_{xxx}=-\sigma_{xyy}
\end{equation}
for the pure-$\beta_1$ model, including the in-plane and out-of-plane Zeeman terms and either of the self-energy treatments used in this work. The modulo-three selection rule underlying this cancellation is the Landau-level manifestation of the cubic harmonicity of the $\beta_1$ SOI.

For completeness, the same Landau-level selection rules determine the remaining tensor components required for the nonlinear-resistivity analysis below. In addition to Eq.~\eqref{eq:Gamma2beta1}, the mixed second-order vertex satisfies
\begin{equation}
\hat\Gamma_{xy}^{\ell-1,\ell+1}=
\begin{pmatrix}
0&0\\
3R_{\beta_1}''\sqrt{\ell+2}&0
\end{pmatrix},
\end{equation}
with the reverse block fixed by Hermiticity. Equivalently, in the off-diagonal block sectors relevant to the trace,
\begin{equation}
\hat\Gamma_{xy}^{r,s}=-\frac{i}{2}(r-s)\hat\Gamma_{xx}^{r,s},
\quad 
r-s=\pm2.
\end{equation}
Combining this identity with Eq.~\eqref{eq:GammayGammaxphase}, the input-index symmetry $\sigma_{ijk}=\sigma_{ikj}$, and the corresponding relation between the current vertices $\hat\Lambda_i=-\hat\Gamma_i$, we obtain
\begin{align}
\sigma_{xxx}&=-\sigma_{xyy}=-\sigma_{yxy}=-\sigma_{yyx},\\
\sigma_{xxy}&=\sigma_{xyx}=\sigma_{yxx}=-\sigma_{yyy}.
\end{align}
Applying the same block-selection argument to the linear Kubo formula in Eq.~(3) of the main text gives
\begin{equation}
\sigma^{(1)}_{xx}=\sigma^{(1)}_{yy},\quad \sigma^{(1)}_{xy}=-\sigma^{(1)}_{yx}.
\end{equation}
These relations hold for the pure-$\beta_1$ model including the in-plane and out-of-plane Zeeman terms and either of the self-energy treatments considered here.

\subsection{Orbital structure of the ladder vertex corrections for white-noise disorder}
\label{sec:vcproof}

For completeness, we show that the exact relation $\sigma_{xxx}=-\sigma_{xyy}$ is preserved by ladder vertex corrections for spin-independent white-noise disorder. We write the dressed vertex in a Green-function channel $a,b=R,A$ as
\begin{equation}
\widetilde{\mathcal V}^{ab}
=
\mathcal V_0+\mathcal L^{ab}\!\left[\widetilde{\mathcal V}^{ab}\right],
\label{eq:BSvertex}
\end{equation}
where $\mathcal L^{ab}$ denotes the impurity ladder kernel. For a scalar disorder potential satisfying
\begin{equation}
\langle U(\bm r)U(\bm r')\rangle=\mathcal W\,\delta(\bm r-\bm r'),
\end{equation}
the orbital part of a single impurity rung contains the Landau-level form factors as
\begin{equation}
{\cal I}_{nn';mm'}\propto\int d^2q\,F_{nm}(\bm q)F_{m'n'}(-\bm q)
\propto\delta_{n-n',\,m-m'}.
\label{eq:vertex_selection}
\end{equation}
Here we used $F_{nm}(\bm q)\propto e^{i(m-n)\phi_{\bm q}}$, so that the last relation follows from the angular integral. Thus, the white-noise ladder kernel preserves the Landau-level-index difference. In the pure-$\beta_1$ block basis $\mathcal B_\ell=(|\ell,\uparrow\rangle,|\ell+3,\downarrow\rangle)^T$, this implies that $\mathcal L^{ab}$ preserves the block-index difference modulo three.

Consequently, dressing cannot mix the different harmonic sectors entering the proof in Sec.~\ref{sec:pert}. The first-order vertices remain in the $r-s=\pm1\pmod3$ sectors, while the second-order vertices remain in the $r-s=0,\pm2\pmod3$ sectors. Moreover, because the scalar white-noise kernel is identical for the $x$ and $y$ vertices and acts independently within each sector, the bare relations in Eqs.~\eqref{eq:GammayGammaxphase} and \eqref{eq:contactminus} are inherited by the dressed vertices:
\begin{align}
\widetilde{\Gamma}^{\,r,s}_y
&=
-i\,\widetilde{\Gamma}^{\,r,s}_x,
\quad r-s=+1\pmod3, \nonumber\\
\widetilde{\Gamma}^{\,r,s}_y
&=
+i\,\widetilde{\Gamma}^{\,r,s}_x,
\quad r-s=-1\pmod3,
\label{eq:dressed_first_vertex}\\
\widetilde{\Gamma}^{\,r,s}_{yy}
&=
-\widetilde{\Gamma}^{\,r,s}_{xx},
\quad r-s=\pm2\pmod3.
\label{eq:dressed_second_vertex}
\end{align}
These relations follow directly from the corresponding Bethe--Salpeter equations: since the ladder kernel preserves each modulo-three sector, the difference between the two sides of Eqs.~\eqref{eq:dressed_first_vertex} and \eqref{eq:dressed_second_vertex} obeys the associated homogeneous ladder equation and therefore vanishes for the regular solution.

The remainder of the proof is then unchanged from Sec.~\ref{sec:pert}. Since the Green functions connect blocks only by $r-s=0\pmod3$, the trace for $\sigma_{xaa}$ closes through the $r-s=\pm2\pmod3$ sector of
\begin{equation}
\widetilde K_a=
\widetilde\Gamma_a G^R\widetilde\Gamma_a+
\widetilde\Gamma_{aa}.
\end{equation}
For the two-first-order-vertex contribution, the two vertices must belong to the same $\pm1$ sector, so Eq.~\eqref{eq:dressed_first_vertex} produces a factor $(\mp i)^2=-1$ between the $y$ and $x$ channels. Equation~\eqref{eq:dressed_second_vertex} gives the same minus sign for the contact term. Hence,
\begin{equation}
\left(\widetilde K_y\right)^{r,s}
=
-\left(\widetilde K_x\right)^{r,s},
\quad r-s=\pm2\pmod3.
\end{equation}
Any dressing of the outer $x$-current vertex is common to $\sigma_{xxx}$ and $\sigma_{xyy}$ and preserves its $r-s=\pm1\pmod3$ sector. Therefore the relative minus sign is unchanged, and we obtain
\begin{equation}
\sigma_{xxx}=-\sigma_{xyy}
\end{equation}
also in the presence of ladder vertex corrections generated by spin-independent white-noise disorder.

\section{Estimate including Joule heating}
\label{sec:heating}

We estimate the experimentally measurable magnitude of the NSdH signal by including Joule heating and the resulting thermal damping of the quantum oscillations. Although the transport coefficients derived in this work are dc quantities, a lock-in measurement employs a slowly varying drive $E(t)=E_0\cos\omega t$, where $E_0$ is the peak electric-field amplitude and $\omega$ is the lock-in angular frequency. We assume throughout that $\hbar\omega$ is much smaller than all relevant electronic energy scales, such that the dc response can be applied quasistatically; the frequency merely labels the $\omega$ and $2\omega$ detection channels.

The Fermi-surface contribution to the second-order conductivity contains the same thermal convolution with $-\partial f/\partial\varepsilon$ as in linear transport~\cite{MP,MN,NLKL2024,Shibata2026}. Namely,
\begin{align}
\sigma_{ijl}^{(2)}(\mu,T)
=
\int d\varepsilon
\left(-\frac{\partial f}{\partial\varepsilon}\right)
{\cal K}_{ijl}^{(2)}(\varepsilon,B_z),
\label{eq:thermal_conv}
\end{align}
where $f=f(\varepsilon-\mu,T)$ is the Fermi--Dirac distribution and ${\cal K}_{ijl}^{(2)}$ is the energy-resolved response kernel. For an oscillatory contribution from a Fermi-surface orbit $a$, ${\cal K}_{ijl}^{(2)}$ contains a factor $\cos[2\pi F_a(\varepsilon)/B_z+\phi_a]$. Using the Onsager relation and linearizing the oscillation frequency around the Fermi level gives
\begin{align}
F_a(\varepsilon)
\simeq
F_a(\mu)
+
\frac{m^*}{|q|\hbar}
(\varepsilon-\mu).
\end{align}
Substitution into Eq.~\eqref{eq:thermal_conv} gives the conventional Lifshitz--Kosevich (LK) factor for the fundamental harmonic~\cite{Ando1982,Shoenberg1984},
\begin{align}
R_T
=
\frac{X}{\sinh X},
\quad
X=
\frac{2\pi^2 k_{\rm B}T m^*}{|q|\hbar |B_z|}.
\label{eq:RT}
\end{align}
Thus, to leading order, the same thermal damping factor applies to linear and nonlinear quantum oscillations associated with the same orbit. This approximation requires the response amplitude, self-energy, and other prefactors to vary slowly over the thermal window $\sim k_{\rm B}T$. We take $m^*=0.0615m_e$; for the present $\beta_1$-only parameters, the SOI-induced difference between the cyclotron masses of the two Fermi surfaces is only a few percent. Furthermore, at the optimal drive amplitudes considered below, $k_{\rm B}T_{\rm e}\simeq0.032$--$0.047$~meV remains smaller than the corresponding cyclotron energy $\hbar\omega_c\simeq0.105$--$0.153$~meV and is comparable to or smaller than the disorder broadening, supporting the use of the LK factor as an order-of-magnitude approximation for the present parameter regime.

We next account for the increase of the effective carrier temperature $T_{\rm e}$ above the lattice temperature $T_{\rm L}$. For strained-Ge 2DHGs, the energy loss to acoustic phonons has been experimentally found to follow~\cite{Drichko2009}
\begin{align}
\frac{P_{\rm e-ph}}{\mathcal A}
=
\Sigma_5
\left(T_{\rm e}^5-T_{\rm L}^5\right),
\label{eq:cooling}
\end{align}
where $P_{\rm e-ph}$ is the power transferred from the two-dimensional hole gas to the lattice and $\mathcal A$ is the active device area. Drichko \textit{et al.} obtained this $T^5$ dependence for $p_{\rm ref}=6.1\times10^{11}$~cm$^{-2}$ and $m_{\rm ref}^*=0.11m_e$, corresponding to an areal cooling coefficient of order $\Sigma_{5,\rm ref}\simeq0.14$~W\,m$^{-2}$K$^{-5}$. In the weak-screening deformation-potential regime, $\Sigma_5\propto(m^*)^2p^{-1/2}$~\cite{Drichko2009,Leturcq2003}. Scaling this value to the present total 2D hole density, $p\simeq2.6\times10^{10}$~cm$^{-2}$, gives $\Sigma_5\simeq0.21$~W\,m$^{-2}$K$^{-5}$. We therefore use the representative value $\Sigma_5=0.20~{\rm W\,m^{-2}K^{-5}}$ below. For the slowly modulated lock-in drive, the cycle-averaged Joule power per area is $\overline P_{\rm J} = \sigma_{xx}^{(1)}E_0^2 / 2$. Balancing this with Eq.~\eqref{eq:cooling} gives
\begin{align}
T_{\rm e}(E_0)
=
\left[
T_{\rm L}^5+
\frac{\sigma_{xx}^{(1)}E_0^2}{2\Sigma_5}
\right]^{1/5}.
\label{eq:Te}
\end{align}
Here we use the cycle-averaged heating as an order-of-magnitude estimate; resolving the thermal dynamics within a lock-in cycle would additionally require the electronic heat capacity and energy-relaxation time. Since our conductivities are two-dimensional sheet conductivities, $[\sigma^{(1)}]={\rm S}$ and $[\sigma^{(2)}]={\rm A\,m/V^2}$. For a Hall bar of width $W$, the peak second-harmonic current is therefore
\begin{align}
I_{2\omega}
=
\frac{W}{2}
\left|\sigma_{xii}^{(2)}\right|
E_0^2
R_T[B_z,T_{\rm e}(E_0)] .
\label{eq:I2heated}
\end{align}
Here, $i=x,y$, and $|\sigma_{xii}^{(2)}|$ denotes the zero-temperature NSdH amplitude at the representative field point considered below. 
We take $W=100~\mu$m, comparable to Hall-bar widths used in Ge-2DHG magnetotransport experiments~\cite{Myronov2023}. If the longitudinal voltage-probe separation is $L$ and the peak drive voltage is $V_0$, then $E_0=V_0/L$. At fixed $E_0$, the local phonon-limited temperature in Eq.~\eqref{eq:Te} is independent of both $W$ and $L$, since both Joule heating and phonon cooling scale with the Hall-bar area $WL$. At fixed $V_0$, however, $L$ enters through $E_0=V_0/L$. In addition, diffusion of heat to cold ohmic contacts introduces a genuine length dependence. Within the Wiedemann--Franz approximation, the corresponding electron--phonon thermal healing length is
\begin{align}
\ell_{\rm ep}
\simeq
\sqrt{
\frac{L_0\sigma_{xx}^{(1)}}
{5\Sigma_5T_{\rm L}^3}
},
\quad
L_0=\frac{\pi^2k_{\rm B}^2}{3e^2}.
\label{eq:lep}
\end{align}
For the parameters below, $\ell_{\rm ep}\simeq80$--$94~\mu$m at $T_{\rm L}=100$~mK. Thus, contact cooling can further reduce $T_{\rm e}$ in sufficiently short devices~\cite{Levitin2022}; we do not include this additional cooling in the estimates below.

Maximizing Eq.~\eqref{eq:I2heated} together with Eq.~\eqref{eq:Te} gives $X\coth X=1+5/[1-(T_{\rm L}/T_{\rm e})^5]$. Since $T_{\rm e}^{\rm opt}\gg T_{\rm L}$ for the present parameters, this reduces to $X_{\rm opt}\simeq6$. We then obtain $T_{\rm e}^{\rm opt}$ and $E_0^{\rm opt}$ self-consistently from Eqs.~\eqref{eq:RT} and \eqref{eq:Te}. Table~\ref{tab:heating_estimate} summarizes the result for four representative points of the $\beta_1$-only calculation in Fig.~1(f). We use $T_{\rm L}=100$~mK, $\Sigma_5=0.20$~W\,m$^{-2}$K$^{-5}$, and $W=100~\mu$m, and optimize the drive amplitude $E_0$ for the thermally damped current in Eq.~\eqref{eq:I2heated}. For this model, $\sigma_{xxx}^{(2)}=-\sigma_{xyy}^{(2)}$, so that their zero-temperature NSdH amplitudes are identical.
\begin{table}[t]
\centering
\begin{tabular}{c c c c c c}
\hline\hline
$B_z^{-1}$ &
$\sigma_{xx}^{(1)}$ &
$|\sigma_{xii}^{(2)}|$ &
$E_0^{\rm opt}$ &
$T_{\rm e}^{\rm opt}$ &
$I_{2\omega}^{\rm max}$
\\
(T$^{-1}$) &
($e^2/h$) &
(nA\,m/V$^2$) &
(V/m) &
(K) &
(fA)
\\ \hline
12.3 & 6.76 & 0.660 & 8.37 & 0.540 & 68.7 \\
14.2 & 7.67 & 0.875 & 5.49 & 0.468 & 39.2 \\
16.0 & 8.55 & 1.043 & 3.86 & 0.415 & 23.1 \\
17.9 & 9.39 & 1.143 & 2.79 & 0.371 & 13.1 \\
\hline\hline
\end{tabular}
\caption{
Estimated NSdH current including Joule heating and LK thermal damping for the pure $\beta_1$ model.
}
\label{tab:heating_estimate}
\end{table}
For example, for $L=200~\mu$m the optimal drive amplitudes in the table correspond to peak voltages of only $0.56$--$1.67$~mV. Although the bare nonlinear conductivity increases toward lower $B_z$, thermal damping becomes progressively stronger, reducing the maximum surviving signal from about $69$ to $13$~fA. Nevertheless, the higher-field points yield second-harmonic currents of several tens of fA, indicating that the predicted NSdH signal lies within a challenging but potentially measurable regime using sensitive low-temperature techniques~\cite{Kuntsevich2015,Tupikov2015}. Substantial efforts are currently being devoted to improving Ge/SiGe material quality and quantum lifetime~\cite{Zhang2024,CSES2025}, as well as to enhancing SOI through heterostructure and strain engineering~\cite{DelVecchio2026,Coppini2026}, with significant recent progress along both directions. These continuing advances are expected to further improve the detectability of the NSdH signal.

The estimate above is intended to establish the relevant signal scale rather than provide a sample-specific prediction. The quantitative thermal suppression depends on sample-specific cooling properties, while additional cooling through the ohmic contacts can improve the thermal conditions. In addition, the current estimate is expressed in the electric-field-driven configuration used in our theoretical formulation; for the current-biased configuration illustrated in Fig.~1(a), the corresponding second-harmonic voltage is obtained through the full conductivity--resistivity conversion discussed in Sec.~\ref{sec:nonlinear_resistivity}.

\section{Nonlinear resistivity} 
\label{sec:nonlinear_resistivity}

The primary experimental observable in a current-driven transport measurement is the nonlinear resistivity rather than the nonlinear conductivity. We therefore discuss their relation without invoking a weak-field approximation. Up to second order, we write the current--electric-field relation and its inverse as
\begin{align}
j_i &= \sigma^{(1)}_{ij}E_j+\sigma^{(2)}_{ijk}E_jE_k+\mathcal{O}(E^3), \label{eq:sigma_rho_relation1}\\
E_i &= \rho^{(1)}_{ij}j_j+\rho^{(2)}_{ijk}j_jj_k+\mathcal{O}(j^3), \label{eq:sigma_rho_relation2}
\end{align}
where ${\rho}^{(1)}=[{\sigma}^{(1)}]^{-1}$. Expanding Eqs.~\eqref{eq:sigma_rho_relation1} and \eqref{eq:sigma_rho_relation2} consistently to second order gives the exact tensor relation~\cite{DLBLK}
\begin{equation}
\rho^{(2)}_{ijk}=-\rho^{(1)}_{ia}\sigma^{(2)}_{abc}\rho^{(1)}_{bj}\rho^{(1)}_{ck}.
\label{eq:rho2_general}
\end{equation}
Importantly, Eq.~\eqref{eq:rho2_general} retains the complete field dependence of the linear resistivity tensor. In the SdH regime, the elements of $\rho^{(1)}_{ij}(B_z)$ can themselves oscillate with magnetic field, and the off-diagonal components are generally finite. Consequently, the detailed oscillation amplitude, beating envelope, and apparent beating-node positions of the directly measured nonlinear resistivity need not coincide exactly with those of the corresponding nonlinear conductivity.

For the pure-$\beta_1$ model, the Landau-level selection rules discussed above impose a particularly simple tensor structure. Defining
\begin{equation}
\sigma_{\rm A}\equiv\sigma_{xxx}=-\sigma_{xyy}=-\sigma_{yxy}=-\sigma_{yyx},
\quad
\sigma_{\rm B}\equiv\sigma_{xxy}=\sigma_{xyx}=\sigma_{yxx}=-\sigma_{yyy},
\label{eq:sigmaAB_full}
\end{equation}
the same selection rules for the linear response give $\sigma^{(1)}_{xx}=\sigma^{(1)}_{yy}$ and $\sigma^{(1)}_{xy}=-\sigma^{(1)}_{yx}$. Hence, the full field-dependent linear resistivity can be written as
\begin{equation}
{\rho}^{(1)}=
\begin{pmatrix}
\rho_{xx} & \rho_{xy}\\
-\rho_{xy} & \rho_{xx}
\end{pmatrix}.
\label{eq:rho1_hallform}
\end{equation}
Substituting Eqs.~\eqref{eq:sigmaAB_full} and \eqref{eq:rho1_hallform} into Eq.~\eqref{eq:rho2_general}, we obtain
\begin{align}
\rho_{xxx}
&=-(\rho_{xx}^2+\rho_{xy}^2)\left(\rho_{xx}\sigma_{\rm A}-\rho_{xy}\sigma_{\rm B}\right), \label{eq:rhoxxx_full}\\
\rho_{xyy}
&=+(\rho_{xx}^2+\rho_{xy}^2)\left(\rho_{xx}\sigma_{\rm A}-\rho_{xy}\sigma_{\rm B}\right), \label{eq:rhoxyy_full}
\end{align}
together with
\begin{align}
\rho_{xxy}=\rho_{xyx}=\rho_{yxx}
&=-(\rho_{xx}^2+\rho_{xy}^2)\left(\rho_{xy}\sigma_{\rm A}+\rho_{xx}\sigma_{\rm B}\right),\\
\rho_{yyy}
&=+(\rho_{xx}^2+\rho_{xy}^2)\left(\rho_{xy}\sigma_{\rm A}+\rho_{xx}\sigma_{\rm B}\right),
\end{align}
and $\rho_{yxy}=\rho_{yyx}=\rho_{xyy}$. Thus, although the field-dependent linear resistivity mixes the two nonlinear-conductivity sectors and can modify the detailed beating pattern of the measured nonlinear resistivity, the characteristic opposite relation of the pure-$\beta_1$ model is preserved exactly:
\begin{equation}
\rho_{xxx}=-\rho_{xyy}.
\label{eq:rho_beta1_exact}
\end{equation}
Therefore, the exact opposite-phase relation between the longitudinal and transverse responses remains directly observable in nonlinear resistivity even at finite $B_z$, without neglecting the quantum oscillations of the linear resistivity or invoking a weak-field approximation. At the same time, Eqs.~\eqref{eq:rhoxxx_full} and \eqref{eq:rhoxyy_full} show that the detailed beating envelope and node positions of $\rho^{(2)}$ can differ from those of $\sigma^{(2)}$ because $\rho_{xx}$, $\rho_{xy}$, and $\sigma_{\rm B}$ are themselves field dependent.

Finally, the tensor conversion is invertible, so the nonlinear conductivity considered throughout the main text can also be reconstructed from experimentally measured transport coefficients. Multiplying Eq.~\eqref{eq:rho2_general} by the linear conductivity tensor gives
\begin{equation}
\sigma^{(2)}_{abc}=-\sigma^{(1)}_{ai}\rho^{(2)}_{ijk}\sigma^{(1)}_{jb}\sigma^{(1)}_{kc}.
\label{eq:sigma2_reconstruction}
\end{equation}
Hence, provided that the relevant components of the linear and nonlinear resistivity tensors are measured, the NSdH conductivity tensor and its phase and beating structure can be reconstructed experimentally. The nonlinear-resistivity measurement therefore provides two complementary levels of information: the exact opposite-phase fingerprint $\rho_{xxx}=-\rho_{xyy}$ of the pure-$\beta_1$ model is directly visible in the measured response, while a full tensor analysis allows direct reconstruction of the nonlinear conductivity for quantitative comparison with the theoretical NSdH spectra.

\end{document}